\documentclass[twocolumn,floatfix,superscriptaddress,longbibliography,PRB]{revtex4}
\usepackage{graphicx,amsfonts,amssymb,amsmath, hyperref}
\usepackage[applemac]{inputenc}
\usepackage{tikz}
\usepackage{amsbsy}
\usepackage{bm}
\usetikzlibrary{arrows,decorations.pathmorphing,backgrounds,positioning,fit,petri,arrows.meta,intersections}
\usepackage{bbold}
\usepackage{bm}
\usepackage{color}
\usepackage{dcolumn}
\usepackage{feynmf} 
\usepackage{graphics}
\usepackage{graphicx}
\usepackage{subfigure}
\usepackage{slashed}
\usepackage{tensor}
\usepackage{placeins}
\usepackage{natbib}
\usepackage{relsize}
\usepackage {ifsym}
\usepackage{diagbox}
\usepackage{blindtext}
\usepackage{tikz}
\usepackage{bm}
\usetikzlibrary{shapes.misc}

\newif\ifhyper
\hypertrue
\ifhyper
\hypersetup{
  citecolor = {green},
  colorlinks = {true}, 
  urlcolor = {blue} 
}
\fi

\newlength{\ldag}
\def\be{\begin{equation}}
\def\ee{\end{equation}}
\def\bea{\begin{eqnarray}}
\def\eea{\end{eqnarray}}
\def\bse{\begin{subequations}}
\def\ese{\end{subequations}}
\def\bc{\begin{center}}
\def\ec{\end{center}}

\newcommand{\comment}[1]{}

\catcode`,\active

\catcode`\,12

\allowdisplaybreaks

\begin{document}

\title{Wrinkling transition in  quenched disordered membranes  at two loops}

\author{O. Coquand} 
\email{oliver.coquand@dlr.de}
\affiliation{Sorbonne Université, CNRS, Laboratoire de Physique Théorique de la Matière Condensée, LPTMC, F-75005 Paris, France}
\affiliation{Institut f\"ur Materialphysik im Weltraum, Deutsches Zentrum f\"ur Luft- und Raumfahrt,
Linder H\"ohe, 51147 K\"oln, Germany}

\author{D. Mouhanna} 
\email{mouhanna@lptmc.jussieu.fr}
\affiliation{Sorbonne Université, CNRS, Laboratoire de Physique Théorique de la Matière Condensée, LPTMC, F-75005 Paris, France}


\begin{abstract}

We  investigate   the flat phase of quenched  disordered  polymerized  membranes  by means of a two-loop, weak-coupling computation  performed near their  upper critical dimension $D_{uc} = 4$, generalizing the one-loop   computation of Morse {\it et al}. [Phys. Rev. A {\bf 45}, R2151 (1992),  Phys. Rev. A {\bf 46}, 1751 (1992)].  Our  work confirms the existence of  the   finite-temperature, finite-disorder,   {\it wrinkling transition},    which has  been recently identified   by Coquand {\it et al}.[Phys. Rev E {\bf 97}, 030102 (2018)] using  a nonperturbative renormalization group approach.   We   also  point  out  ambiguities  in  the   two-loop computation  that  prevent  the  exact   identification of   the properties  of the novel  fixed point associated  with  the wrinkling transition, which  very likely requires a three-loop order approach.

\end{abstract}

\maketitle

{\sl Introduction}

The  critical and, more generally,  the long-distance  equilibrium statistical physics of pure, homogeneous,   systems is now  widely  understood. By  contrast,  quenched,   random heterogeneities,  such as  defects or impurities,  inevitably present in most   real physical  systems,  are known to give  rise to a  wide spectrum of new phenomena.  Quenched disordered membranes  occupy   a special  place;  see e.g. \cite{radzihovsky04},  as their most famous physical realizations  seem to bring out  the  physical effects  of both random {\it bonds} and random {\it fields};  see  \cite{dotsenko01,young98,nattermann98,dedominicis98,dedominicis06,wiese07,tarjus20} for reviews.  For instance,   in a  series of experiments  beginning in the early 90's,  Mutz {\it et al}. \cite{mutz91}  then  Chaieb {\it et al}.  \cite{chaieb06,chaieb08}    have shown that, upon cooling  below  the  chain melting temperature,    photo-induced partially polymerized  vesicles made of diacetylenic phospholipid undergo  a   transition from a smooth structure, at high polymerization,  to a {\it wrinkled} structure, at low  polymerization,   with randomly frozen normals that could  characterize  a {\it glassy phase}.  This transition  and the resulting wrinkled phase   have   been  conjectured to  result from  the joint effects of random heterogeneities  on  both  the internal metric and  the curvature of the membrane \cite{radzihovsky91b}  that  bear  formal similarities  with, respectively,  random bonds  and  random fields   in magnetic systems;  see below.   More recently,  in the context of the rapidly growing    {\it defect engineering}   \cite{banhart11,liu15,yang18,ni18} of graphene \cite{novoselov04,novoselov05},  it has been shown that by thoroughly damaging a clean  sheet of this material  with a laser beam, it was   possible  to  induce a crystal-to-glass  transition   giving rise to a  vacancy-amorphized  graphene  structure \cite{kotakoski11,eder14,joo17}. Here  also,  the  inclusion of lattice defects -- foreign adatoms or substitutional impurities --   is expected to lead, in addition to metric alterations, to a rearrangement of sp$^2$-hybridized carbon atoms into nonhexagonal structures and, thus,  to the  generation of nonvanishing curvature structures showing again that the underlying physics could  rely on  the coexistence of the two kinds of disorder.

Disordered membranes  also  stand   out from the crowd   by   the   theoretical investigations to which they have been subjected.   Stimulated by the work of Mutz {\it et al}. \cite{mutz91}  on partially polymerized vesicles,   the first   attempt  to probe the effects of   disorder in membranes  has been realized   by Nelson and Radzihovsky \cite{nelson91,radzihovsky91b},   who have  focused their study on the role of disorder in the preferred metric.  Performing a weak-coupling  expansion near the upper critical dimension  $D_{\hbox{\scriptsize uc}}=4$ they  have shown that, for $D<4$,  while the disorder is irrelevant and the renormalization group (RG) flow is driven toward   the disorder-free fixed point -- $P_4$ -- at any finite temperature, an instability close to $T=0$ could be accompanied by a diverging Edwards-Anderson correlation length,  leading  to a glassy phase.   Then  Radzihovsky and Le Doussal \cite{radzihovsky92},  by   employing a large embedding dimension $d$-expansion,   have confirmed  such a possibility, finding  an  instability  of the flat phase   toward   a spin-glass-like phase.   However no  quantitative  or qualitative  empirical  confirmation of  this scenario has  been provided.  Morse {\it et al.}  \cite{morse92a,morse92b} have  then reconsidered the weak-coupling analysis of Ref.\cite{nelson91,radzihovsky91b} within an approach including  both  metric  and curvature  disorders.  They have  confirmed the irrelevance of the disorder  in $D<4$  but  shown that  the presence of  curvature disorder  gives rise  to a new and {\it vanishing temperature} fixed point -- $P_5$ --  {\it stable} with respect to randomness but  {\it unstable} with respect to the temperature.   Somewhat disappointing from the point of view of the search for   a new exotic phase, these works  have been followed  mainly by mean-field approximations involving either  short-range  \cite{radzihovsky92,bensimon92,bensimon93,attal93,park96,mori96,benyoussef98} or   long-range   \cite{ledoussal93,mori94a}   disorder  leading to speculate about   the existence of a glassy phase  at any temperature and for large enough disorder;  see \cite{ledoussal18} for a review.   Again no confirmation of  this conjecture has been provided. However very recently,  an approach based on the nonperturbative renormalization group (NPRG), following those performed on disorder-free membranes \cite{kownacki09,braghin10,essafi11,hasselmann11,essafi14,coquand16a} has been realized by Coquand {\it et al}. \cite{coquand18}  on the model  considered by Morse and Lubensky displaying  both metric and curvature disorders.   Their  main result   has been  to identify  a novel   {\it finite temperature, finite disorder}   fixed-point --  $P_c$ -- once  {\it unstable}, and   thus  associated with  a  second-order phase transition  and making  the $T=0$ fixed point {\it fully    attractive} provided $T<T_c$.   This study  has allowed   the identification of    three distinct  universal scaling behaviors  \cite{coquand20}  corresponding  both qualitatively {\it  and} quantitatively to  those observed in the experiments of Chaieb {\it et al} \cite{chaieb06,chaieb08}.

Whereas the NPRG  results have  been  challenged  within a recent self-consistent screening approximation  approach \cite{ledoussal18}  they have been   confirmed within a large $d$ approach  performed  at next-to-leading order in $1/d$ \cite{saykin20b}  although in a model involving only curvature disorder.   In this controversial context,  it was compelling  to further investigate  the   model of Morse {\it et al.}    \cite{morse92a,morse92b}. In that  respect,   an essential   feature of the novel fixed point  $P_c$  found in  \cite{coquand18}  is that  its coordinates near $D_{\hbox{\scriptsize uc}}=4$   differ  only from those of the  vanishing temperature fixed point $P_5$  by  terms   of order $\epsilon^2$,  with $\epsilon=4-D$, strongly suggesting  that  $P_c$ could be  also identified   within a  perturbative  $\epsilon$-expansion  up to this  order. This is the reason why we   investigate, in this  letter,  quenched disordered membranes at  two-loops  in the vicinity of the upper critical  dimension,  extending both the one-loop computation  of Morse {\it et al.}    performed  30 years  ago \cite{morse92a,morse92b},  at the next order   and the recent  two-loop  computation  of  Coquand {\it et al.}  \cite{coquand20a}  -- see also \cite{mauri20} --   on  disorder-free membranes,  to the disordered case.   We  derive    the RG equations,  analyze them  and provide the critical quantities, notably the anomalous dimension $\eta$,  at order $\epsilon^2$.    Our approach confirms unambiguously the existence of  the  once-unstable  fixed point  $P_c$ characterizing  a phase transition between a high-temperature  phase  controlled by the disorder-free fixed point $P_4$ and a low-temperature phase controlled by the vanishing-temperature, finite-disorder,   fixed point $P_5$ \footnote{Note that we  consider  here the {\it flat phase} of membranes; also  the  {\it  "high-temperature"}  phase discussed here should not  be confused with the crumpled phase of membranes.}.   It nevertheless  reveals also  a drawback   of  the perturbative approach at two-loop   order  which manifests  through  the impossibility to identify exactly  the location and properties of the fixed points $P_5$  and $P_c$ at this order;  we  argue  that  this    should very likely  be raised by a three-loop order computation.

\vspace{0.2cm} 

{\sl The action}

\vspace{0.2cm}

A membrane is  modelized by a   $D$-dimensional surface embedded in  a    $d$-dimensional Euclidean space. A  point on   the membrane is thus  identified  by   $D$-dimensional  vector  ${\bf x}$ and  a configuration of the membrane in the Euclidean space is described  through the embedding  ${\bf x}\to {\bf R} ({\bf x})$ with ${\bf R}\in  \mathbb{R}^d$.     In the flat phase we define   the average position of a point  ${\bf x}$:  
\begin{equation}
{\bf {R}}^0({\bf x})=[\langle {\bf R}({\bf x})\rangle]=\zeta  x_i \bm{ e}_i 
\label{flat}
\end{equation}
where the $\bm{ e}_i$ form  an orthonormal set of $D$ vectors and $\zeta$ is the stretching factor taken to be  one in what follows.  In Eq.(\ref{flat}) $[\dots]$  and $\langle\dots\rangle$ denote  averages  over  disorder  and   thermal fluctuations respectively. 
The fluctuations around the configuration (\ref{flat})  are parametrized  by writing  $\bf{ R}(\bm{x)}={\bf {R}}^0({\bf x}) + {\bf u}({\bf x}) + {\bf h}({\bf x})$  with ${\bf h}.\bm{ e}_i=0$. The fields     ${\bf u}$ and ${\bf h}$ represent $D$ longitudinal  -- phonon -- and   $d-D$ transverse -- flexural  -- modes, respectively.   The  long-distance, effective,  action   is  given by   \cite{morse92a,morse92b}: 
\begin{equation}
\begin{array}{ll}
{\cal S}=\displaystyle \int \text{d}^Dx & \hspace{-0.2cm} \displaystyle \ \Big\{{\kappa \over 2}\big({\Delta} {\bm h} ({\bf x})\big)^2 
+ {\lambda\over 2}\,  u_{ii}({\bf x})^2 +  {\mu}\,  u_{ij}({\bf x})^2 \\
& -{\bm c}({\bf x}).{\Delta} {\bm h} ({\bf x})  - \sigma_{ij}({\bf x}) \, u_{ij}(\bf{ x})\Big\}\ .
\label{action}
\end{array}
\end{equation}
In Eq.(\ref{action}) the first term represents the   curvature  energy    with bending rigidity  $\kappa$ while  the second and third terms  represent the elastic energies  with  $u_{ij}$ being  the strain tensor  which,  truncated to its most relevant part reads 
$u_{ij}\simeq  {1\over 2} \left[\partial_i u_j+\partial_i  u_j+ \partial_i   {\bm h} . \partial_j   {\bm h}  \right]$;  $\lambda$ and $\mu$ are  the Lamé coefficients;  The  fourth and fifth terms in Eq.(\ref{action}) represent disorder fields   ${\bm c}$ and  $\sigma_{ij}$  that couple  respectively to the  curvature $\Delta{\bm h}$  --  thus linearly to  ${\bm h}$ as a random field \footnote{with the major difference that  ${\bm c}$ couples  with  the derivative of  the order parameter  ${\Delta} {\bm h}$    and not directly to the order parameter $\partial_i   {\bm h}$.}  --  and  to the strain tensor $u_{ij}$ -- thus  quadratically  to  ${\bm h}$ as a random mass.  These fields  are chosen  to be short-range  quenched Gaussian ones  with  zero-mean value and variances given by  \cite{morse92a,morse92b}: 
\begin{equation}
\begin{array}{ll}
[c_{i}({\bf  x})\  c_{j}({\bf x}')]= \Delta_{\kappa}\,  \delta _{ij} \ \delta^{(D)}({\bf x-x'})
\\
\\
\big[\sigma_{ij}({\bf x})\   \sigma_{kl}({\bf x'})\big]= (\Delta_{\lambda} \delta _{ij} \delta_{kl}+2 \Delta_{\mu} I_{ijkl})\ \delta^{(D)}({\bf x-x'})
\label{variance}
\end{array}
\end{equation}
where $I_{ijkl}={1\over 2}(\delta_{ik}\delta_{jl}+\delta_{il}\delta_{jk})$, with $i,j,k,l=1\dots D$.   Stability considerations require that the coupling constants  $\kappa$,  $\mu$,  and $\lambda+(2/D) \mu$    as well as    $\Delta_{\kappa}$,  $\Delta_{\mu}$ and $\Delta_{\lambda}+(2/D) \Delta_{\mu}$ are  positive.

Disorder averages are  performed through  the replica trick which leads to the effective  action  \cite{morse92a,morse92b}:
\begin{equation}
\begin{array}{ll}
 \hspace{-0.2cm}S =\displaystyle  \int    \text{d}^Dx \  \bigg\{& \displaystyle \hspace{-0.2cm}  {\widetilde Z_{\alpha\beta} \over 2} {\Delta}\bm{h}^{\alpha} ({\bf x}){\Delta}\bm{h}^{\beta} ({\bf x}) + \displaystyle    {\widetilde\lambda_{\alpha\beta}\over 2}  u_{ii}^{\alpha}({\bf x})  u_{jj}^{\beta}({\bf x}) 
\label{S}
\\
\hspace{-0.9cm}  &  \displaystyle   + {\widetilde \mu_{\alpha\beta}}\,   u_{ij}^{\alpha}({\bf x})  u_{ij}^{\beta}({\bf x})   \bigg\} 
\end{array}
\end{equation}
where Greek indices are associated with   the $n$ replica. In  Eq.(\ref{S})  we   have  rescaled the fields  $\bm{h}\mapsto  T^{1/2} Z^{1/2} \kappa^{-1/2} \bm{h}$, $\bm{u}\mapsto  T  Z  \kappa^{-1} \bm{u}$ where $Z$ is a   field renormalization   and  introduced  the running coupling constants:    $\widetilde \lambda=\lambda TZ^2 \kappa^{-2}$,  $\widetilde \mu_k=\mu TZ^2 \kappa^{-2}$,   $\widetilde\Delta_{\lambda}=\Delta_\lambda Z^2\kappa^{-2}$, $\widetilde\Delta_{\mu}=\Delta_\mu Z^2\kappa^{-2}$,  $\widetilde\Delta_{\kappa}=\Delta_\kappa T^{-1}Z\kappa^{-1} $ and   $\widetilde Z^{\alpha\beta} =Z \,\delta^{\alpha\beta} -\widetilde \Delta_\kappa  \,J^{\alpha\beta}$, $\widetilde\mu^{\alpha\beta} = \widetilde\mu\,\delta^{\alpha\beta} - \widetilde\Delta_\mu     \,J^{\alpha\beta}$ and 
$\widetilde\lambda^{\alpha\beta} = \widetilde\lambda \,\delta^{\alpha\beta} - \widetilde\Delta_\lambda \,J^{\alpha\beta}$ where $J^{\alpha\beta}\equiv 1$ $\forall\,  \alpha,\beta$. Note that  $\widetilde \mu$  and  $\widetilde \lambda$   can be used as  a  measure  of the temperature $T$   while $\widetilde\Delta_{\kappa }$ diverges at vanishing temperatures.  Finally, as usual, on defines the correlation functions   $G_{{h_i}{h_j}}({\bm q})= \big[\langle {h_i}({\bm q}) h_j(-{\bm q})\rangle \big]$ and $G_{{u_i}{u_j}}({\bm q})= \big[\langle {u_i}({\bm q}) u_j(-{\bm q})\rangle \big]$  as well  the thermal  -- $\chi({\bm q})$ -- and disorder --  $C({\bm q})$ -- ones through \cite{morse92a,morse92b}:
\begin{equation}
\begin{array}{ll}
G_{{h_i h_j}}({\bm q})&= \big[\langle \delta {h_i}({\bm q})  \delta h_j(-{\bm q})\rangle \big]+\big[\langle {h_i}({\bm q})\rangle \langle h_j(-{\bm q})\rangle \big]\\ 
\\
&= T \chi_{{h_i h_j}}({\bm q})+ C_{{h_i h_j}}({\bm q})
\label{correlationh}
\end{array}
\end{equation}
and
\begin{equation}
\begin{array}{ll}
G_{{u_i u_j}}({\bm q})&= \big[\langle \delta {u_i}({\bm q})  \delta u_j(-{\bm q})\rangle \big]+\big[\langle {u_i}({\bm q})\rangle \langle u_j(-{\bm q})\rangle \big]\\ 
\\
\label{correlation}
&= T \chi_{{u_i u_j}}({\bm q})+ C_{{u_i u_j}}({\bm q})
\end{array}
\end{equation}
with $\delta h_i({\bm q})=h_i({\bm q})-\langle h_i({\bm q}) \rangle$,  $\delta u_i({\bm q})=u_i({\bm q})-\langle u_i({\bm q}) \rangle$.  At low momenta one expects the scaling behaviors \cite{morse92a,morse92b}:  
\begin{equation}
\begin{array}{ll}
&\chi_{{h_i h_j}}({\bm q})\sim q^{-(4-\eta)} \  \, , \hspace{0.3cm}   C_{{h_i h_j}}({\bm q})\sim  q^{-(4-\eta')}
\\
&\chi_{{u_i u_j}}({\bm q})\sim q^{-(4-\eta_u)}, \hspace{0.3cm}   C_{{u_i u_j}}({\bm q})\sim  q^{-(4-\eta_u')}\ . 
\end{array}
\end{equation}
Ward identities relate these quantities \cite{morse92a,morse92b} through  $\eta_u  + 2\eta  = 4-D$ and  $\eta'_u + 2\eta' = 4-D$.  Finally  one   defines \cite{morse92a,morse92b,ledoussal18}, from  $\eta$ and $\eta'$,  the   exponent  $\phi=\eta'-\eta$  that  determines  which kind  of   -- thermal or disorder -- fluctuations dominates  at a given fixed point:  (i) if $\phi>0$, the fixed point behavior is dominated by thermal fluctuations   (ii)  if $\phi<0$  the fixed point behavior is dominated by  disorder fluctuations  (iii)  if $\phi=0$ both fluctuations coexist; the fixed point is said to be marginal.

\vspace{0.3cm} 

{\sl  Renormalization group equations and fixed points}
		
\vspace{0.3cm} 

As in the disorder-free  \cite{guitter88,guitter89,coquand20a} case Ward identities associated with   a partial rotation invariance ensure the renormalizability of the theory.   Also  only the renormalizations of phonon and flexural mode propagators are required. As in  \cite{coquand20a} we    have  treated  the massless theory  using the  modified minimal substraction  scheme   and  used   standard techniques for computing  massless Feynman diagram calculations; see,  e.g., \cite{Kotikov:2018wxe}.  
As usual one defines  dimensionless coupling  constants   $\overline\mu=Z^{-2}\, k^{D-4}\widetilde\mu$,     $\overline\lambda=Z^{-2}\, k^{D-4}\widetilde\lambda$,   $\overline\Delta_{\mu}=Z^{-2}\, k^{D-4}\widetilde\Delta_{\mu}$,  $\overline\Delta_{\lambda}=Z^{-2}\, k^{D-4}\widetilde\Delta_{\lambda}$ and  $\overline\Delta_{\kappa}=Z^{-1}\, \widetilde\Delta_{\kappa}$.   The  running anomalous dimension is given by $\eta_t=\partial_t \ln Z$  and $\phi_t=\eta_t'  - \eta_t=  \partial_t\ln \overline\Delta_{\kappa}$   \footnote{We indicate  a misprint in \cite{coquand18} where this relation was incorrectly written $\eta_t'  = \eta_t +  \partial_t\ln \widetilde\Delta_{\kappa}$.} where $t=\ln \overline k$, $\overline k$ being a  renormalization momentum scale \footnote{Related to $k$ by $\overline{k}^2=4\pi e^{-\gamma_E}k^2$ where $\gamma_E$ is the Euler constant.}.   The RG  equations  are given  in Appendix A and  computational  details will be given in a forthcoming publication \cite{coquand20c}.  Note finally that our computations have been checked using the effective-field theory obtained by integrating over the Gaussian  phonon-field  ${\bf u}$  \cite{nelson87,radzihovsky92,ledoussal18,coquand20c}, see below.

 Let us first recall the one-loop results \cite{morse92a,morse92b}. At this order  one finds, in $D<4$,  two nontrivial  physical fixed points,    located on the hypersurfaces $\overline\lambda/\overline\mu=\overline\Delta_{\lambda}/\overline\Delta_{\mu}=-1/3$. First is   the disorder-free  fixed point,  $P_4$, for which  $\overline\mu=96\pi^2\,\epsilon/(24+d_c)$, $\overline\Delta_{\mu}=\overline\Delta_{\kappa}=0$  and  $\eta=\eta'/2=\phi=12\,\epsilon/(24+d_c)$. It is   fully attractive  and thus controls the  long distance behaviour of both disordered and disorder-free membranes.  This fixed point is -- obviously -- dominated by  thermal fluctuations.  There is another fixed point,   $P_5$,  located at vanishing temperature, i.e. $\overline\mu=0$. To get  this fixed point  from the RG equations one has to consider  the coupling constant $\overline g_{\mu}=\overline\mu\, \overline\Delta_{\kappa}$  that stays  finite at vanishing temperature while $\overline\Delta_{\kappa}$ is diverging. $P_5$ is characterized by  $\overline\Delta_\mu=24\pi^2\,\epsilon/d_{c6}$,  $\overline g_\mu=48\pi^2\,\epsilon/d_{c6}$ and $\eta=\eta'=3\,\epsilon/d_{c6}$ with $d_{c6}=d_c+6$.  At this fixed point  one has  $\phi=0$;  it is thus marginal.  A further analysis taking account of nonlinear contributions  shows that $P_5$ is marginally  {\it relevant} \cite{morse92a,morse92b}.

At two-loop  order we   recover  the disorder-free fixed point  $P_4$ whose   coordinates and anomalous dimension have been given in \cite{coquand20a}.   Using the variables  relevant to study  the vanishing temperature we also identify a fixed point with  $\overline\mu=0$ that coincides  with the fixed point $P_5$ found at one-loop order. Note however  that  we   are   not able to  fully characterize this  fixed point -- see below.   Finally the search for a new fixed point is inspired by the NPRG results \cite{coquand18}  where we  recall  that the  coordinates of $P_c$  in the vicinity of $D_{uc} = 4$  are  given at leading nontrivial order in $\epsilon$   by  \cite{coquand18,coquand20}:  $\overline\mu=4\pi^2\epsilon^2(5d_c +27)/15 d_{c6}^2+ O(\epsilon^3)$, $\overline\lambda=-1/3\,  \overline\mu+O(\epsilon^3)$, $\overline\Delta_{\mu}=24\pi^2\epsilon/d_{c6}+O(\epsilon^2)$, $\overline\Delta_{\lambda}=-1/3\, \overline\Delta_{\mu}+O(\epsilon^2)$ and $\overline\Delta_{\kappa}= 180d_{c6}/(27+5 d_c)\epsilon$ while the anomalous dimension  is given by: 
\begin{equation}
\displaystyle \eta_c^{\hbox{\scriptsize nprg}}=\frac{3 \,\epsilon}{d_{c6}}-\frac{d_c(425\, d_c+2556)}{240\, d_{c6}^3}\epsilon^2 
\label{etacnprg}
\end{equation}
with $\eta_c'^{\hbox{\scriptsize nprg}}=\eta_c^{\hbox{\scriptsize nprg}}$.  Within the perturbative context we  thus consider, for  the various coupling constants,   the ansatz: 
\begin{equation}
\overline{X} = \mathcal{C}_X^{(1)}\,\epsilon + \mathcal{C}_X^{(2)} \epsilon^2\hspace{0.3cm}  \hbox{for}  \hspace{0.3cm} \overline{X} =\{\overline\lambda, \overline\mu, \overline\Delta_\lambda , \overline\Delta_\mu \}
\label{coup1}
\end{equation}
 where the $\mathcal{C}_X^{(1)}$ are given by the coordinates of the vanishing temperature fixed point  $P_5$ at one-loop order \footnote{whose coordinates only differ from those of $P_c$   by terms of order $\epsilon^2$.},  and the unusual -- singular -- behaviour  for $ \overline\Delta_{\kappa}$: 
\begin{equation}
  \overline\Delta_{\kappa}= {C_{\Delta_{\kappa}}^{(-1)}/ \epsilon}+ C_{\Delta_{\kappa}}^{(0)}\ . 
\label{coup2}
\end{equation}
Canceling the RG  equations at  (next-to-leading) order  $\epsilon^3$ for  the $\overline{X}$'s and  at  order  $\epsilon$  for  $\overline\Delta_{\kappa}$ we   find a new fixed point $P^*$  with parameters:
\begin{equation}
\begin{split}
& \mathcal{C}_\lambda^{(2)} = - \frac{\mathcal{C}_\mu^{(2)}}{3} \,, \\[0.2cm]
& \mathcal{C}_{\Delta_\lambda}^{(2)} =\frac{\mathcal{C}_\mu^{(2)}d_c}{6\,  d_{c6}}- \frac{2(6 d_c+ 83)\pi^2}{5\, d_{c6}^3} , \\[0.2cm]
&\mathcal{C}_{\Delta_\mu}^{(2)} = - \frac{\mathcal{C}_\mu^{(2)} d_c}{2\, d_{c6}}- \frac{6(14 d_c + 37) \pi^2}{5\, d_{c6}^3} , \\[0.2cm]
& \mathcal{C}_{\Delta_\kappa}^{(0)} = -\frac{2\, {(d_c+3)}}{d_{c6}} -\frac{4 \big(28 \,d_c +27\big)\pi^2}{5\, \mathcal{C}_\mu^{(2)}d_{c6}^3},\\
&\mathcal{C}_{\Delta_\kappa}^{(-1)} = \frac{48 \pi^2}{\mathcal{C}_\mu^{(2)}{d_{c6}}}\ . 
\label{parameters}
\end{split}
\end{equation} 

As seen in these expressions one of the parameters entering in (\ref{coup1})-(\ref{parameters}), here  $\mathcal{C}_\mu^{(2)}$,   is left {\it  undetermined}.  An analysis of the NPRG  approach \cite{coquand18}   shows  that   using   the ansatz  (\ref{coup1})-(\ref{coup2})  and   canceling the  corresponding  NPRG   equations  at the same order  in $\epsilon$ leads to the same difficulty, i.e. the same indetermination of $\mathcal{C}_\mu^{(2)}$,  which is thus a feature of the $\epsilon$-expansion  and not of the loop expansion.   It is thus  judicious to go beyond the former expansion.  One can first   analyze   the RG equations numerically.  Doing this we  clearly identify a once-unstable  fixed point   in the vicinity of $D=4$ with coordinates  of the type  (\ref{coup1})-(\ref{parameters}).  Thereafter, in order to identify analytically this fixed point  one  can  push  the solution of the RG equations beyond  next-to-leading order, notably by  canceling  the  equation  for $\overline\Delta_{\kappa}$
at order  $\epsilon^2$.   This raises the indetermination on  $\mathcal{C}_\mu^{(2)}$ which is found to be equal to:
\begin{equation}
C_{\mu, \hbox{\scriptsize 2f}}^{(2)} =\displaystyle  \frac{4 \pi ^2 \left(3075\,  d_c^2+16850 \, d_c-576\right)}{15\, d_{c6}^2 \left(166 + 169\, d_c+ 20\, d_c^2\right)}
\label{Cmu}
\end{equation}
where the index $\hbox{\scriptsize 2f}$ refers to the two-field  (phonon-flexuron)  theory. 
 Note that this value is approximate as one expects a three-loop contribution  to  (\ref{Cmu}). 
However with  the  expressions  (\ref{parameters}) and (\ref{Cmu})   we  reproduce very satisfactorily  the numerical results  in the extreme vicinity of $D=4$, e.g.  for $\epsilon$ of order  $10^{-3}$  the errors for the coordinates are at worst of  order  $10^{-8}$.  
 We now give  the eigenvalues around $P^*$ at leading non-vanishing order:   
\begin{equation}
\left\{\frac{3\, d_c\, \mathcal{C}_{\mu}^{(2)}}{8\; d_{c6}^3}\, \epsilon^2\: ;\ 
-\frac{d_c}{d_{c6}}\,   \epsilon\: ;   -\frac{d_c}{d_{c6}}\, \epsilon\: ;\  -\epsilon\: ;\  -\epsilon\right\}\
\nonumber
\end{equation}
with $\mathcal{C}_{\mu}^{(2)}$ given by (\ref{Cmu}) which is positive for any physical value of $d_c$. 
Having one repulsive direction  the fixed point $P^*$   is associated with  a second order phase transition. It is  characterized by  the anomalous dimensions: 
\begin{equation}
\displaystyle\  \eta_c^{\hbox{\scriptsize 2l}} = \frac{3 \epsilon}{d_{c6}} - \frac{d_c\big(5\,\mathcal{C}_\mu^{(2)}d_{c6}^2 + 2(407+60d_c)\pi^2\big)}{80\pi^2 d_{c6}^3}\epsilon^2\  
\label{anomalous} 
\end{equation}
and    $\eta_c'=\eta_c$ implying   $\phi=0$ so   that $P^*$  is marginal. The result (\ref{anomalous}) is also  found within the effective (pure flexuron)  approach of the theory, see \cite{coquand20c} and Appendix B, which is a strong confirmation of the validity of our result. Note however that  in the  latter case the approximate expression of  $\mathcal{C}_\mu^{(2)}$  slightly  differs and is given by:
\begin{equation}
C_{\mu,\hbox{\scriptsize eff}}^{(2)} =\displaystyle  \frac{4 \pi ^2 \left(3450\,  d_c^2+ 19100 \, d_c-576\right)}{15\, d_{c6}^2 \left(166 + 169\, d_c+ 20\, d_c^2\right)}\ . 
\label{Cmueff}
\end{equation}
However  this change affects extremely weakly the physical results -- see below. 

\begin{figure}[h!]
\includegraphics[scale=0.45]{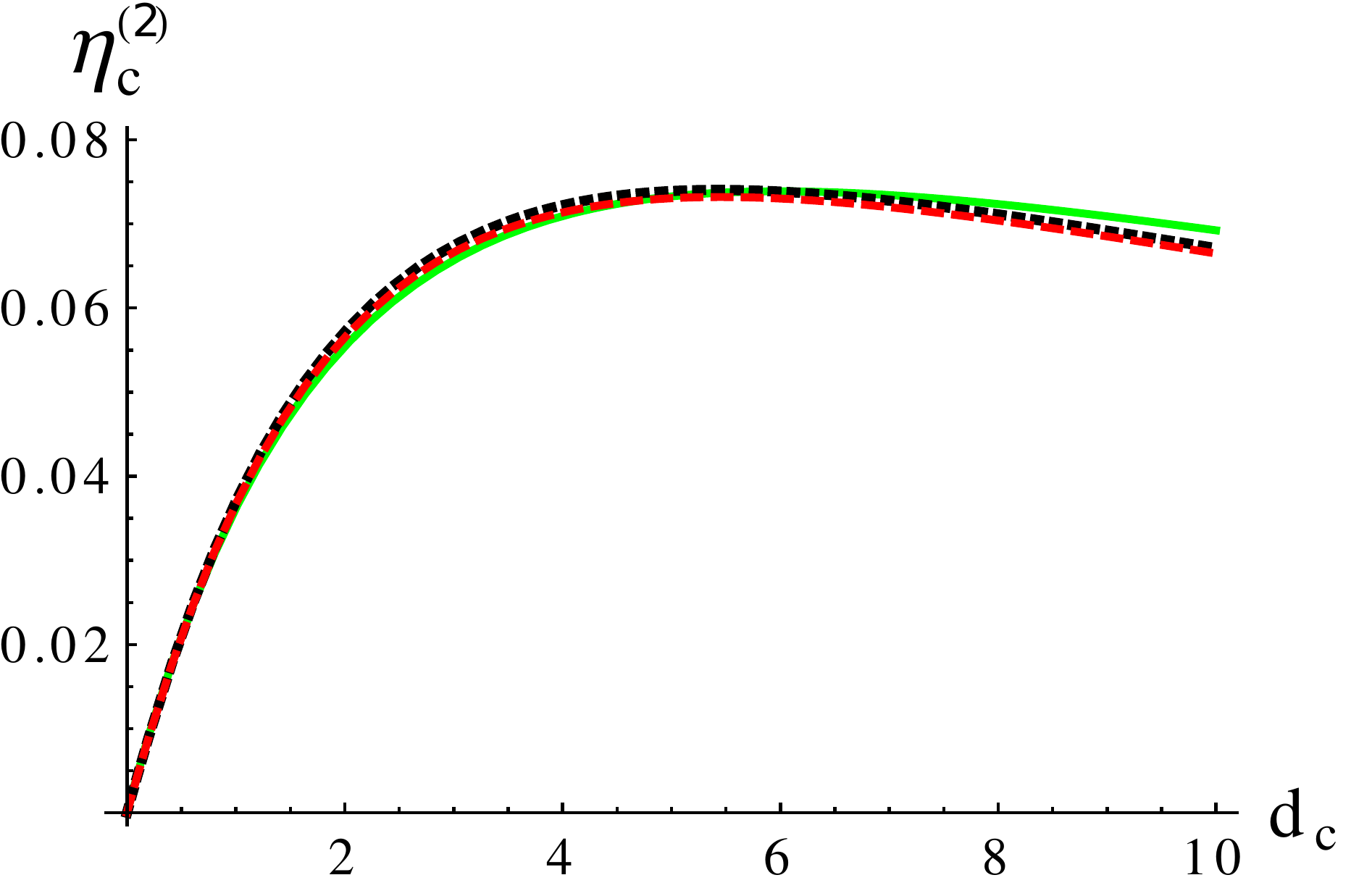}
\caption{The correction of  $O(\epsilon^2)$  to $\eta_c$, $\eta_c^{\hbox{\scriptsize(2)}}$,  as function of $d_c$ at the  fixed point $P_c$. The solid  line shows  the prediction from the NPRG approach \cite{coquand18}; the  dashed  line shows   the two-loop/two-field result (present work); the  dotted  line shows  the two-loop/efffective result (present work).}
\label{etaPC}
\end{figure}

All the qualitative properties of $P^*$ -- one  marginally relevant direction of order $\epsilon^2$ and  one coupling constant $\mu$ of order $\epsilon^2$ -- are shared with those of the fixed point $P_c$ found in  \cite{coquand18,coquand20} using a  NPRG approach. Moreover the  agreement  between the anomalous dimension obtained within the present work,  {\it i.e.}  (\ref{anomalous})   with  $\mathcal{C}_{\mu}^{(2)}$ given by (\ref{Cmu}) or (\ref{Cmueff}),  and that computed  with the NPRG approach (\ref{etacnprg})   is remarkable,  see Fig.(\ref{etaPC}) where we  have  represented the two-loop corrections $\eta_c^{(2)}$ defined as $\eta_c=\eta_c^{(1)}+\eta_c^{(2)} \epsilon^2$ as functions of $d_c$. In the  physical situation  -- $d_c=1$  --  they  are given by:   $ \eta_c^{\hbox{\scriptsize (2)NPRG}}=0.0362$,     $\eta_{c,\hbox{\scriptsize 2f}}^{\hbox{\scriptsize(2)2l}}=0.0366$ and $\eta_{c,\hbox{\scriptsize eff}}^{\hbox{\scriptsize(2)2l}}=0.0370$.  We thus identify  $P^*$ with $P_c$ and fully confirms the existence of a -- wrinkling -- phase transition at finite temperature.

Concerning the fixed point $P_5$ we  find -- numerically -- that it is, in fact marginally irrelevant  -- in agreement with the unstable character of $P_c$ and with the  NPRG approach. However as said above there are difficulties to characterize  $P_5$ --  as well as $P_c$ -- at low-temperatures. Indeed   this    implies to use the  low-temperature variables $\overline g_{\lambda}=\overline\lambda\, \overline\Delta_{\kappa}$ and   $\overline g_{\mu}=\overline\mu\, \overline\Delta_{\kappa}$ at order $\epsilon^2$.  But, in the vicinity  of $P_5$  -- or $P_c$ --  one   has   at next-to-leading order in $\epsilon$:
 $$\overline{g}_\mu =  \mathcal{C}_\mu^{(2)}\mathcal{C}_{\Delta_\kappa}^{(-1)} \epsilon +
 \mathcal{C}_\mu^{(3)}\mathcal{C}_{\Delta_\kappa}^{(-1)} \epsilon^2 + \mathcal{C}_\mu^{(2)}\mathcal{C}_{\Delta_\kappa}^{(0)} \epsilon^2 +O(\epsilon^3)$$
 where it appears  that, due to the specific scaling of  $\overline\Delta_{\kappa}$  with $\epsilon$ that involves negative powers of this parameter,  the   subsubleading contribution  in $\epsilon$  to $\mu$ -- $\mathcal{C}_\mu^{(3)}$ -- is  needed but is obviously  lacking within the present   --  two-loop   order -- computation.

\vspace{0.3cm} 

{\sl Conclusion}
		
\vspace{0.3cm}

We  have  investigated  quenched disordered  membranes by means of  a two-loop order  perturbative approach.  As a main result our approach clearly confirms the  finding obtained with the NPRG approach \cite{coquand18}, i.e.  the existence of a richer phase diagram than that  expected from  previous  investigations:  the existence of a novel fixed point $P_c$ characterizing a {\it wrinkling} phase transition occurring at a temperature   $T_c$  separating a   disorder-free phase at $T>T_c$ controlled by the  vanishing-disorder attractive fixed point $P_4$ and a low-temperature $T<T_c$ ``glassy-phase" controlled by the vanishing-temperature, finite-disorder,   attractive fixed point $P_5$.   We  thus have  reached  a consistent  picture of  disordered membranes at finite temperatures  and in particular of the occurrence of a  wrinkling transition.  Our  work  reinforces  the  interest  to investigate experimentally or numerically   this  transition   in several systems  involving both curvature and stretching  disorder.  This  includes  (i)  a further  study of  partially polymerized fluid vesicles that have been already  investigated by  Chaieb {\it et al}.  and have shown to be qualitatively and quantitatively well  explained by the  scenario  proposed in \cite{coquand18,coquand20}  and  (ii)  a careful investigation of graphene and  graphene-like materials   with quenched lattice defects.   Moreover our work, by  confirming   the  attractive character of the   vanishing temperature fixed point $P_5$,    opens   the possibility of a  low-temperature phase  controlled by a complex energy  landscape and a genuine  ``glassy phase" that have been  intensively looked for theoretically. It is  thus pressing to probe  this phase  experimentally and numerically, notably in the context of the physics  of graphene where it would be  of  prime interest to study  the effects induced by disorder on the  electronic and transport properties of  graphene and graphene-like materials in this phase.  Finally, from a more formal point of view our work   strongly  suggests  to investigate  deeply the nature of the perturbative series  in the vicinity of the fixed points $P_5$ and  $P_c$.  In particular it would be of interest, even if it would  represent a very substantial amount of work,   to  see whether the three-loop contributions  indeed raise the ambiguities encountered within  the two-loop order computation when studying the wrinkling transition.

\vspace{3cm}

\acknowledgements

We   wish  to greatly  thank  S. Teber  for   discussions.


\appendix
\section{Renormalization group equations: two-field  theory}

One  gives  here the  two-loop  RG  equations  for the -- dimensionless -- coupling constants (forgetting their overlining)   entering in action Eq.(\ref{S}) :

\begin{widetext}
\vspace{-0.5cm}
\begin{equation}
\begin{split}
                                                      \partial_t{\mu}&=-{\mu}\,\eta_{ut}+\frac{d_c\,{\mu}^2}{6(16\,\pi^2)}(1+2{\Delta}_\kappa)\\
						&+\frac{d_c\,{\mu}^2\big({\mu}(686{\Delta}_\kappa^2+908{\Delta}_\kappa+227)({\lambda}
						+{\mu})({\lambda}+2{\mu})-227{\Delta}_\lambda{\mu}^2
						(4{\Delta}_\kappa+1)-227{\Delta}_\mu(4{\Delta}_\kappa+1)({\lambda}^2+2{\lambda}{\mu}
						+2{\mu}^2)\big)}{216(16\,\pi^2)^2\,({\lambda}+2{\mu})^2}\,,
						\nonumber
					\end{split}
				\end{equation}
				
				\vspace{-0.05cm}
				\begin{equation}
					\begin{split}
						  \hspace{-1cm}\partial_t{\lambda} &= -{\lambda}\,\eta_{ut} + \frac{d_c\,(6{\lambda}^2+6{\lambda}{\mu}
						+{\mu}^2)}{6(16\,\pi^2)}(1+2{\Delta}_\kappa)
						+\frac{d_c}{216(16\,\pi^2)^2\,({\lambda}+2{\mu})^2}\bigg[{\mu}({\lambda}+2{\mu})
						\Big(-2{\Delta}_\kappa^2\big(108(d_c+9){\lambda}^3\\
						&+ 18(16d_c+63){\lambda}^2{\mu}+(156d_c+239){\lambda}{\mu}^2+(24d_c+77){\mu}^3\big)
						-4{\Delta}_\kappa \big(54(d_c+7){\lambda}^3+72(2d_c+3){\lambda}^2{\mu}\\
						&+ (78d_c-179){\lambda}{\mu}^2+(12d_c-17){\mu}^3\big) -\big(54(d_c+7){\lambda}^3
						+72(2d_c+3){\lambda}^2{\mu}+(78d_c-179){\lambda}{\mu}^2+(12d_c-17){\mu}^3\big)\Big)\\
						& +{\Delta}_\lambda{\mu}^2(4{\Delta}_\kappa+1)(378{\lambda}^2-162{\lambda}{\mu}
						-17{\mu}^2)+{\Delta}_\mu (4{\Delta}_\kappa+1)(378{\lambda}^4+594{\lambda}^3{\mu}
						+415{\lambda}^2{\mu}^2	-358{\lambda}{\mu}^3-34{\mu}^4)\bigg]\,,
							\nonumber
					\end{split}
				\end{equation}

                              \begin{equation}
                               \hspace{-1.5cm}
					\begin{split}
					\partial_t{\Delta}_\mu &= -{\Delta}_\mu\,\eta_{ut}+\frac{d_c\,{\mu}\,\big(2{\Delta}_\mu
						(2{\Delta}_\kappa+1)-{\Delta}_\kappa^2{\mu}\big)}{6(16\,\pi^2)}
						+\frac{d_c\, {\mu}}{108 \, (16\pi^2)^2\,({\lambda}+2{\mu})^2}\bigg[{\Delta}_\lambda
						\big(343{\Delta}_\kappa^2 {\mu}^3 -227{\Delta}_\mu {\mu}^2(4{\Delta}_\kappa+1)\big)\\
						& +{\Delta}_\mu {\mu} \big(343{\Delta}_\kappa^2 (3{\lambda}^2+8{\lambda}{\mu}
						+6{\mu}^2) +908{\Delta}_\kappa ({\lambda}+{\mu})({\lambda}+2{\mu}) +227({\lambda}
						+{\mu})({\lambda}+2{\mu})\big)\\
						& -{\Delta}_\kappa^2 {\mu}^2 (232{\Delta}_\kappa+227)({\lambda}+{\mu})
						({\lambda}+2{\mu}) -227{\Delta}_\mu^2 (4{\Delta}_\kappa+1)({\lambda}^2+2{\lambda}
						{\mu}+2{\mu}^2)\bigg]\,,
				\nonumber
					\end{split}
				\end{equation}
			
				\begin{equation}
				 \hspace{-0.5cm}
					\begin{split}
						\partial_t{\Delta}_\lambda &= -{\Delta}_\lambda\,\eta_{ut}-\frac{d_c\,{\Delta}_\kappa^2\,
						(6{\lambda}^2+6{\lambda}{\mu}+{\mu}^2)}{6(16\,\pi^2)}
						+ \frac{2 d_c\big(3{\Delta}_\lambda (2{\lambda} +{\mu}) +{\Delta}_\mu
						(3{\lambda}+{\mu})\big)}{6(16\,\pi^2)}(1+2{\Delta}_\kappa)\\
						&-\frac{d_c}{108 \, (16\pi^2)^2 ({\lambda}+2{\mu})^2}\bigg[{\Delta}_\kappa^2\Big({\Delta}_\lambda
						{\mu}\big(216(d_c+9){\lambda}^3+18(52d_c+387){\lambda}^2{\mu}	+72(16d_c+63){\lambda}{\mu}^2
						+(288d_c+401){\mu}^3\big)\\
						&\: +{\Delta}_\mu\big(108(d_c+9){\lambda}^4+36(16d_c+63){\lambda}^3{\mu}+3(348d_c+995)
						{\lambda}^2{\mu}^2+16(45d_c+79){\lambda}{\mu}^3+6(24d_c+77){\mu}^4\big)\Big)\\
						&\: +(4{\Delta}_\kappa+1)({\lambda}+2{\mu})\Big(9d_c(3{\lambda}+{\mu})({\lambda}
						+2{\mu})\big(2{\Delta}_\lambda{\mu}+{\Delta}_\mu({\lambda}+{\mu})\big)
						+{\mu}({\lambda}+{\mu})\big({\Delta}_\lambda(378{\lambda}-81{\mu})-{\Delta}_\mu
						(81{\lambda}+17{\mu})\big)\Big)\\
						&\: +{\Delta}_\kappa^2{\mu}({\lambda}+2{\mu})\Big(-4{\Delta}_\kappa\big(27(d_c+11)
						{\lambda}^3+9(8d_c+51){\lambda}^2{\mu}+(39d_c+209){\lambda}{\mu}^2+(6d_c+47){\mu}^3\big)\\
						&\: -\big(54(d_c+7){\lambda}^3+72(2d_c+3){\lambda}^2{\mu}+(78d_c-179){\lambda}{\mu}^2
						+(12d_c-17){\mu}^3\big)\Big)\\
						&\: -(4{\Delta}_\kappa+1)\big({\Delta}_\lambda(378{\lambda}-81{\mu})-{\Delta}_\mu(81{\lambda}
						+17{\mu})\big)\big({\Delta}_\lambda{\mu}^2+{\Delta}_\mu({\lambda}^2+2{\lambda}{\mu}
						+2{\mu}^2)\big)\bigg]\,,
							\nonumber
					\end{split}
				\end{equation}

				\begin{equation}
				 \hspace{-2cm}
					\begin{split}
						 \partial_t{\Delta}_\kappa &= {\Delta}_\kappa\,\eta_t- \frac{5\,{\Delta}_\kappa\big({\Delta}_\lambda
						{\mu}^2+{\Delta}_\mu({\lambda}^2+2{\lambda}{\mu}+2{\mu}^2)\big)}{16\,\pi^2\,({\lambda}
						+2{\mu})^2}\\
						&+\frac{{\Delta}_\kappa}{36 (16\,\pi^2)^2\,({\lambda}+2{\mu})^4}\bigg[-{\Delta}_\kappa^2{\mu}^2
						({\lambda}+2{\mu})^2\big((39d_c+490){\lambda}^2+4(39d_c+110){\lambda}{\mu}+(81d_c+130){\mu}^2\big)\\
						&+{\mu}({\lambda}+2{\mu})\bigg(5{\Delta}_\lambda{\mu}^2\big(3(15d_c+47){\Delta}_\kappa
						{\mu}+(15d_c+32){\mu}+411{\Delta}_\kappa{\lambda}+122{\lambda}\big)\\
						&+{\Delta}_\mu\Big({\lambda}^3\big(3(39d_c+415){\Delta}_\kappa+(39d_c+340)\big)+3{\lambda}^2{\mu}
						\big(234d_c{\Delta}_\kappa+78d_c+435{\Delta}_\kappa+70\big)\\
						& +2{\lambda}{\mu}^2\big((477d_c+600){\Delta}_\kappa+(159d_c+50)\big)+2{\mu}^3\big(3(81d_c+55)
						{\Delta}_\kappa+(81d_c-20)\big)\Big)\bigg)\\
						& + 30{\mu}^2\big(-53{\Delta}_\lambda^2{\mu}^2-{\Delta}_\lambda{\Delta}_\mu(61{\lambda}^2
						+32{\lambda}{\mu}+32{\mu}^2)\big)
						 -30{\Delta\mu}^2\big(
						17{\lambda}^4+14{\lambda}^3{\mu}+10{\lambda}^2{\mu}^2-8{\lambda}{\mu}^3-4{\mu}^4\big)\bigg]\,,
							\nonumber
					\end{split}
				\end{equation}
				
with 				
			 \vspace{-0.2cm}			
			\begin{equation}
			 \hspace{-0.7cm}
				\begin{split}
					\eta_t & = \frac{5{\mu}\big({\lambda}+{\mu})}{16\,\pi^2({\lambda}+2{\mu})}
					-  \frac{5\Big(2{\Delta}_\mu ({\lambda}^2+2{\lambda}{\mu}+2{\mu}^2)
					-{\mu}\big({\Delta}_\kappa({\lambda}^2+3{\lambda}{\mu}+2{\mu}^2)
					-2{\Delta}_\lambda{\mu}\big)\Big)}{16\,\pi^2\,({\lambda}+2{\mu})^2} \\
					& +\frac{{\mu}^2 \big((-39\,d_c-340){\lambda}^2-4(39\,d_c+35){\lambda}{\mu}+(20-81\,d_c){\mu}^2\big)}
					{72 (16\,\pi^2)^2\,({\lambda}+2{\mu})^2}\\
					&-\frac{1}{72 (16\,\pi^2)^2\,({\lambda}+2{\mu})^4}\bigg[2 ({\lambda}+2{\mu})\bigg({\mu}
					(2{\Delta}_\kappa+1)\Big({\Delta}_\mu \big((-39d_c-340){\lambda}^3-6(39d_c+35){\lambda}^2{\mu}
					-2(159d_c+50){\lambda}{\mu}^2\\
					&+2(20-81d_c){\mu}^3\big)-5{\Delta}_\lambda{\mu}^2\big((15d_c+32){\mu}+122{\lambda}\big)\Big)
					-2880\pi^2({\lambda}+2{\mu})\big({\Delta}_\lambda{\mu}^2+{\Delta}_\mu({\lambda}^2
					+2{\lambda}{\mu}+2{\mu}^2)\big)\bigg)\\
					& +3{\Delta}_\kappa{\mu}^2({\lambda}+2{\mu})^2\Big({\lambda}^2\big(39(d_c+10){\Delta}_\kappa
					+39d_c+340\big)+4{\lambda}{\mu}\big((39d_c+60){\Delta}_\kappa+39d_c+35\big) \\
					&+{\mu}^2\big((81d_c+30){\Delta}_\kappa	+ 81d_c-20\big)\Big)+20\Big(53{\Delta}_\lambda^2{\mu}^4
					+{\Delta}_\lambda{\Delta}_\mu{\mu}^2(61{\lambda}^2+32{\lambda}{\mu}+32{\mu}^2)\\
					& +{\Delta}_\mu^2(17{\lambda}^4+14{\lambda}^3{\mu}+10{\lambda}^2{\mu}^2
					-8{\lambda}{\mu}^3-4{\mu}^4)\Big)\bigg]
						\nonumber
				\end{split}
			\end{equation}
	
	and  $\eta_{ut}  = \epsilon-2\eta_t\,. $

		\end{widetext}

\section{Renormalization group equations:  effective field theory}

The effective field theory is obtained after an integration over the phonon field $\mathbf{u}$ in Eq.(\ref{S}), see \cite{coquand20a}:
	\begin{equation}
		\begin{split}
			& S_{\hbox{\scriptsize eff}} = \int_{\mathbf{k}}\frac{\widetilde{Z}_{\alpha\beta}}{2}k^4\, \mathbf{h}^\alpha(\mathbf{k}).
			\mathbf{h}^\beta(-\mathbf{k}) \\
			& +{1\over 4} \int_{\mathbf{k}_1,\mathbf{k}_2,\mathbf{k}_3,\mathbf{k}_4}\!\!\!\!\!\!\!\!\!\!\!\!\!\!\!\!\!\!\!\!\!\!
			\mathbf{h}^\alpha(\mathbf{k}_1).\mathbf{h}^\alpha(\mathbf{k}_2)
			\widetilde{R}^{ab,cd}_{\alpha\beta}(\mathbf{q}) k_1^a\, k_2^b\, k_3^c\, k_4^d\,
			\mathbf{h}^\beta(\mathbf{k}_3).\mathbf{h}^\beta(\mathbf{k}_4)
		\end{split}
			\nonumber
	\end{equation}
	where $\int_{\mathbf{k}} = \int {d^Dk}/(2\pi^D)$, $\mathbf{q} = \mathbf{k}_1 + \mathbf{k}_2 = -\mathbf{k}_3 - \mathbf{k}_4$, and the interaction tensor $\widetilde{R}^{ab,cd}_{\alpha\beta}$ is defined as follows:
	\begin{equation}
		\widetilde{R}^{ab,cd}_{\alpha\beta}(\mathbf{q}) = \widetilde{b}^{\alpha\beta} N_{ab,cd}(\mathbf{q})
		+ \widetilde{\mu}^{\alpha\beta} M_{ab,cd}(\mathbf{q})\, .
	\end{equation}

In this expression  the transverse tensors $N$ and $M$ are defined as  a function of the projector transverse to  $\mathbf{q}$, $P^T_{ab} = \delta_{ab} - q_aq_b/{\bf q}^2$, by:

	\vspace{2cm}
	\begin{equation}
		\begin{split}
			& N_{ab,cd}(\mathbf{q}) = \frac{1}{D-1} P^T_{ab}(\mathbf{q}) P^T_{cd}(\mathbf{q}) \\
			& M_{ab,cd}(\mathbf{q}) = \frac{1}{2}\big[P^T_{ac}(\mathbf{q}) P^T_{bd}(\mathbf{q}) + 
			P^T_{ad}(\mathbf{q}) P^T_{bc}(\mathbf{q})\big] - N_{ab,cd}(\mathbf{q}) \, . 
		\end{split}
			\nonumber
	\end{equation}
	
	\vspace{-0.5cm}
	
	Finally the coupling constant $\widetilde{b}_{\alpha\beta} = \widetilde{b}\, \delta_{\alpha\beta} - \widetilde{\Delta}_b\,
	J_{\alpha\beta}$ is related to the bare couplings $\widetilde{\mu}$, $\widetilde{\lambda}$, $\widetilde{\Delta}_\mu$ and
	$\widetilde{\Delta}_\lambda$ by:
	
	\vspace{1cm}
	\begin{equation}
		\begin{split}
			& \widetilde{b} = \frac{\widetilde{\mu}(D\widetilde{\lambda} + 2\widetilde{\mu})}
			{\widetilde{\lambda} + 2 \widetilde{\mu}} \\
			& \widetilde{\Delta}_b = \frac{D(\widetilde{\Delta}_\mu\widetilde{\lambda}^2 + 2\widetilde{\Delta}_\lambda
			\widetilde{\mu}^2)- 2\widetilde{\mu}\big(\widetilde{\Delta}_\lambda \widetilde{\mu}
			- 2 \widetilde{\Delta}_\mu(\widetilde{\lambda} + \widetilde{\mu})\big)}
			{(\widetilde{\lambda} + 2 \widetilde{\mu})^2} \,.
		\end{split}
			\nonumber
	\end{equation}
	
\vspace{1cm}

\begin{widetext}
	The two-loop  RG  equations are then given by:

	\begin{equation}
	 \hspace{-1cm}
		\begin{split}
			\partial_t\mu &= - \mu\, \eta_{ut} + \frac{d_c\,  \mu^2}{6(16\pi^2)}(1+2\Delta_\kappa) \\
			&+\frac{d_c\,\mu^2}{1296(16\pi^2)^2}\bigg[ \mu(574 +2296 \Delta_\kappa +1732 \Delta_\kappa^2)
			+b(107 +428 \Delta_\kappa +326 \Delta_\kappa^2) -(574\Delta_\mu+107\Delta_b)(1+4\Delta_\kappa)\bigg]\,,
		\end{split}
			\nonumber
	\end{equation}

	\begin{equation}
		 \hspace{-1.7cm}
		\begin{split}
		\partial_tb &= - b\, \eta_{ut} + \frac{5\, d_c\,  b^2}{12(16\pi^2)}(1+2\Delta_\kappa) \\
			&-\frac{5\, d_c\, b^2}{2592(16\pi^2)^2}\bigg[(178\Delta_\mu-91\Delta_b)(1+4\Delta_\kappa)
			+b(91 +364 \Delta_\kappa +394\Delta_\kappa^2) -2\mu(89 +356\Delta_\kappa +146\Delta_\kappa^2)\bigg]\,,
		\end{split}
			\nonumber
	\end{equation}

\begin{equation}
 \hspace{-2.2cm}
		\begin{split}
			 \partial_t\Delta_\mu &= - \Delta_\mu\, \eta_{ut} -  \frac{d_c\, \mu\big(\mu\Delta_\kappa^2 -
			2\Delta_\mu(1+2\Delta_\kappa)\big)}{6(16\pi^2)}\\
			 &-\frac{d_c\,\mu}{648(16\pi^2)^2} \bigg[107\Delta_b\Delta_\mu(1+4\Delta_\kappa)
			-b\Delta_\mu(107 + 428\Delta_\kappa + 326\Delta_\kappa^2) + 574\Delta_\mu^2(1+4\Delta_\kappa)
			-163 \mu\, \Delta_b\, \Delta_\kappa^2 \\
			&+ b\, \mu\, \Delta_\kappa^2(107+112\Delta_\kappa) -2\mu\Delta_\mu (287+1148\Delta_\kappa+ 1299\Delta_\kappa^2)
			+574\Delta_\kappa^2\mu^2 +584\Delta_\kappa^3\mu^2\bigg]\,,
		\end{split}
			\nonumber
	\end{equation}

	\begin{equation}
	 \hspace{-2.5cm} 
		\begin{split}
			 \partial_t\Delta_b &= - \Delta_b\, \eta_{ut} -  \frac{5\, d_c\, b\big(b\Delta_\kappa^2 -
			2\Delta_b(1+2\Delta_\kappa)\big)}{12(16\pi^2)}\\
			& -\frac{5d_c\,b}{1296(16\pi^2)^2} \bigg[b^2\Delta_\kappa^2(91 + 212\Delta_\kappa)
			- b\, \Delta_b(91+364\Delta_\kappa +591\Delta_\kappa^2)+2b\, \Delta_\kappa^2 \big(73\Delta_\mu
			+\mu(32\Delta_\kappa - 89)\big) \\
			&+\Delta_b\big( (91\Delta_b - 178\Delta_\mu)(1+4\Delta_\kappa) + 2\mu(89 + 356\Delta_\kappa
			+146\Delta_\kappa^2) \big)\bigg]\,,
		\end{split}
			\nonumber
	\end{equation}

 \begin{equation}
  \hspace{-4.5cm}
		\begin{split}
 \partial_t\Delta_\kappa &= \Delta_\kappa\,  \eta_{t} - \frac{5(\Delta_b + 2\Delta_\mu)}{6(16\pi^2)}\Delta_\kappa\\
			&+\frac{\Delta_\kappa}{1296(16\pi^2)^2}\bigg[5b^2(15d_c - 242)\Delta_\kappa^2 -2\Big(795\Delta_b^2
			-870\Delta_b\Delta_\mu -60\Delta_\mu^2 +5\Delta_b\mu(58+129\Delta_\kappa) \\
			&- 2\mu\, \Delta\mu\big(111d_c(1+3\Delta_\kappa) +5(33\Delta_\kappa -4) \big) + 2\mu^2\Delta_\kappa^2
			(130 + 111d_c)\Big ) \\
			&-5b\Big( \Delta_b\big(15d_c(1+3\Delta_\kappa) -212 - 681\Delta_\kappa\big)
			+2\big(\Delta_\mu(58+129\Delta_\kappa) - 56\mu\Delta_\kappa^2\big)\Big)\bigg]\,,
		\end{split}
			\nonumber
	\end{equation}

with
	
	\begin{equation}
	\hspace{-1cm}
		\begin{split}
			\eta_t &= \frac{5(b+2\mu)}{6(16\pi^2)}(1+\Delta_\kappa) - \frac{5(\Delta_b + 2\Delta_\mu)}{6(16\pi^2)} \\
			&+\frac{1}{2592(16\pi^2)^2}\bigg[-1060\Delta_b^2 + 5b^2\big(15d_c(1+3\Delta_\kappa+3\Delta_\kappa^2) 
			-2(106+318\Delta_\kappa + 333\Delta_\kappa^2)\big) + 80\Delta_\mu^2 \\
			&+8\mu\Delta_\mu(1+2\Delta_\kappa)(111d_c-20) - 4\mu^2\big(111d_c(1+3\Delta_\kappa+3\Delta_\kappa^2)
			+10(9\Delta_\kappa^2-6\Delta_\kappa-2)\big)\\
			&+1160\Delta_b\big(\Delta_\mu - (1+2\Delta_\kappa)\mu\big)
			-10b\big(\Delta_b(15d_c-212)(1+2\Delta_\kappa) +116\Delta_\mu(1+2\Delta_\kappa)
			-4\mu(29 + 87\Delta_\kappa + 72\Delta_\kappa^2)\big)\bigg]
		\end{split}
	\nonumber
	\end{equation}

	and  $\eta_{ut}  = \epsilon-2\eta_t\,. $

	\end{widetext}

\end{document}